\documentclass[a4paper,11pt]{article}

\usepackage[latin1]{inputenc}
 \usepackage[T1]{fontenc}
 \usepackage[normalem]{ulem}
 \usepackage[english]{babel}
 \usepackage{verbatim}
 \usepackage{graphicx}
\usepackage{algorithmic} 
\usepackage{algorithm} 
\usepackage[latin1]{inputenc} 
\usepackage{multirow}
\usepackage[square, comma, sort&compress]{natbib}
\usepackage[margin=1.2in]{geometry}
\usepackage{chngpage}
\usepackage[font=small,format=plain,labelfont=bf,up,textfont=it,up]{caption}

\usepackage{amssymb} 
\usepackage{mathabx} 

\begin{document}

\title{Quick overview of inference methods in PLoM: \\{\it combining fast and exact plug-and-play algorithms\\ to achieve flexibility, precision and efficiency} }
\author{ ANTOINE {BASSET}$^1$, JOSEPH {DUREAU}$^{1,2*}$, \\SEBASTIEN {BALLESTEROS}$^{3*}$, BERNARD {CAZELLES}$^1$\\
{\it {\footnotesize 1: UMR 7625, UPMC-CNRS-ENS, Ecole Normale Superieure, Paris, France{}}}\\
{\it {\footnotesize 2: Statistics Department, London School of Economics, London, United Kingdom{}}}\\
{\it {\footnotesize 3: Department of Ecology and Evolutionary Biology, Princeton University, United States{}}}\\
{\it {\footnotesize * to whom correspondance should be addressed}}}

\maketitle

\begin{figure}[!h]
\begin{center}
\includegraphics[width=10cm,trim=4 4 4 60, clip=true]{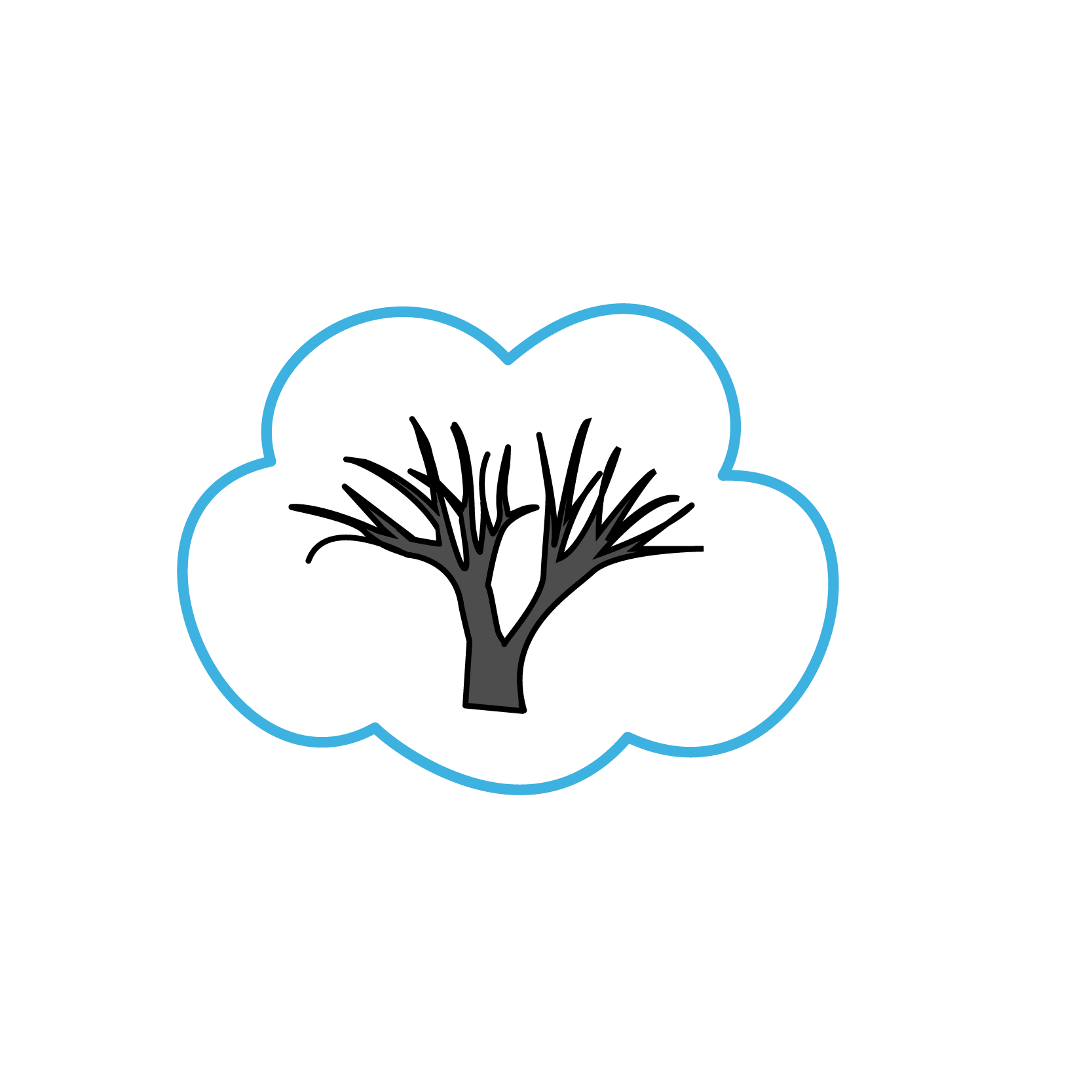}
\end{center}
\end{figure}
\thispagestyle{empty}
\setcounter{page}{-1}

\pagebreak{}
\thispagestyle{empty}

\topskip0pt
\vspace*{\fill}
\tableofcontents
\vspace*{\fill}

\pagebreak{}

\section{Introduction: aiming for flexibility, precision and efficiency}

The process of fitting and validating dynamic models against epidemic and outbreak data is often labor and computationally intensive. The net result of this is that scientists and public health agencies often use suboptimal methods to explain outbreak and epidemic data. Nevertheless, there has been intense scientific activity around what is called \emph{plug-and-play} methods: flexible algorithms that can be applied with hardly any effort to new models, by just plugging a short model-specific bit of code into them.

This effort has lead to a rich diversity of algorithms that are ready to be used. Some are very quick to run, because they rely on some clever mathematical or numerical approximations, and some others offer the possibility to make no approximation whatsoever (aside from the model itself, of course), but they can quickly become prohibitively complex and computationally expensive to run. The goal of PLoM is to take the best out of this variety of methods, making it easier and more feasible to use the most precise and demanding algorithms, while keeping everything plug-and-play.

In addition, by taking advantage of code templating approaches and symbolic computation libraries, PLoM brings an extra layer of simplicity: you won't have to code a single line to plug your model and start playing with the PLoM library.

\begin{figure}[!h]
\begin{center}
\includegraphics[width=13cm,trim=4 25 4 4, clip=true]{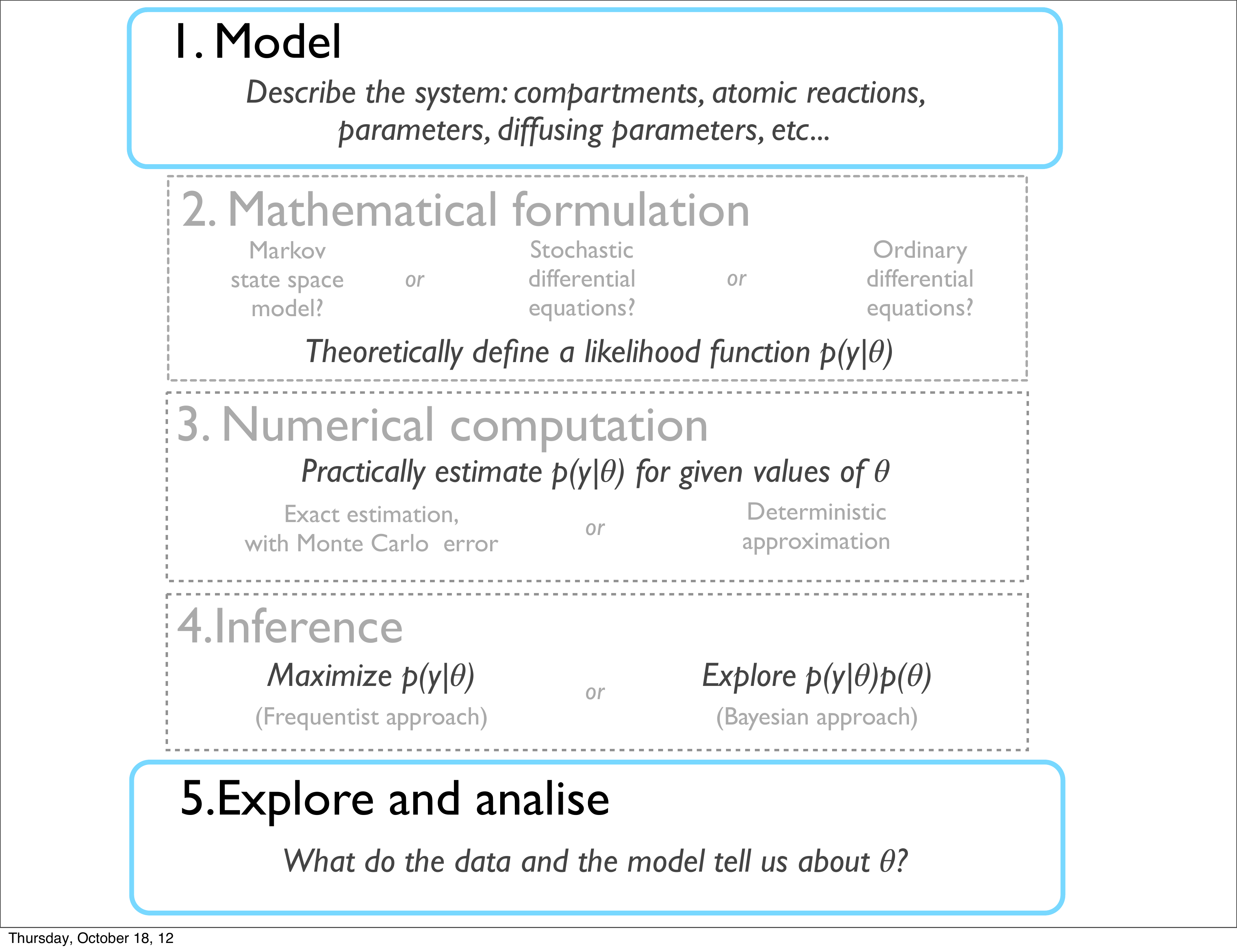}
\caption{\label{fig:Layers}Layered representation of a classic workflow in epidemiology. PLoM lets you concentrate on steps 1 and 5: model, explore, and analyse.}
\end{center}
\end{figure}

Figure \ref{fig:Layers} illustrates the classic workflow of mathematical modellers. The plug-and-play paradigm aims at allowing you to concentrate on the first and fifth layers: modelling, exploration, and analysis. This document will hopefully give all you need to understand about the three intermediary layers in order to take the best out of the library of methods provided by PLoM.

 The first intermediary layer is the mathematical formulation of the model. PLoM integrates the principal approaches that can be found in the literature to translate compartmental epidemic models  into mathematical terms. These formulations lead to different ways of computing the implied likelihood, which is the second intermediary layer.  Deterministic models can simply be integrated using Euler schemes, nonlinear stochastic differential equations can be integrated (with a certain level of approximation) with tools as the Extended Kalman Filter (EKF), and in some cases no tractable exact or approximated scheme is available. This is the case of some more general stochastic Markovian state space models that can only be sampled from (i.e. simulated, one scenario at a time). In such cases, the likelihood is estimated by propagating hundreds or thousands of scenarios, or particles, through the model. Lastly, the third intermediary layer deals with extracting information from the data (noted $y$) regarding the underlying parameters of the model (contained in a vector noted  $\theta$). This can either be done by maximising the likelihood $p(y|\theta)$, which is the frequentist way of doing inference, or by going Bayesian and exploring the posterior density $p(\theta|y)\propto p(y|\theta)p(\theta)$ that includes the a priori beliefs regarding $\theta$, reflected by $p(\theta)$.

As we just mentioned, there is a family of methods to estimate $p(y|\theta)$ that work for all Markovian state space models, which naturally includes ODE's and SDE's. These methods are called particle filters (see \cite{Doucet2009} for more details). Basically, they can be used with any Markovian compartmental model, whatever its mathematical formulation.  Not only are these filters very generic, but they also provide an unbiased (also called exact) estimate of the likelihood. Nevertheless, this estimate has some variability, termed Monte Carlo error, that can be brought to 0 by having the number of particles tend to infinity. In practice, hundreds or thousands of particles are generally needed to obtain acceptable levels of variability, which makes particle filters computationally expensive, as well as their associated inference algorithms: iterated filtering (MIF, see \cite{Ionides2011}) and particle MCMC (pMCMC, see \cite{Andrieu2010}). What we propose is to make these methods more accessible by using faster (but approximate) algorithms to initialise them. Indeed, initialisation phases can be critically long, both for optimization problems or when running an MCMC: it is just practical sense to let faster methods do their part of the work! 

An overview of the algorithms implemented in PLoM is presented in Table  \ref{tab:algs}. In the next section, we will illustrate on a simple example how the \texttt{lhs}, \texttt{ksimplex} and \texttt{kmcmc} algorithms can be used to literally shorten the calibration phase of an adaptive particle MCMC.

 \pagebreak{}

\topskip0pt
\vspace*{\fill}

\begin{table}[h]
\begin{adjustwidth}{-5in}{-5in}
\begin{centering}
\begin{tabular}{|c|c|c|c|c|c|}
\hline 
\multirow{2}{*}{\textbf{Goal}} & \textbf{Inference}  & \textbf{Numerical}  & \textbf{Mathematical}  & \textbf{Cost (order of magnitude)} & \textbf{PLoM}  \tabularnewline
 & \textbf{algorithm} & \textbf{resolution} & \textbf{formulations}& \textbf{per iteration} & \textbf{command} \tabularnewline
\cline{1-6} 
& Iterated & \multirow{2}{*}{Particle filter$^1$} & \multirow{2}{*}{Markov models} & \multirow{2}{*}{$O(1000\times k\times n)$ } & \multirow{2}{*}{\texttt{mi}f} \tabularnewline
  &  Filtering &  &  &  & \tabularnewline
\cline{2-6} 
Maximize & \multirow{4}{*}{Simplex} & \multirow{2}{*}{ODE integration$^2$} & \multirow{2}{*}{ODE}& \multirow{2}{*}{$O(d^{2}\times k\times n)$ } & \multirow{2}{*}{\texttt{simplex}} \tabularnewline
 $p(y|\theta)$&  &  &  &  & \tabularnewline
\cline{3-6} 
  (or $p(\theta|y)$)&  & SDE integration$^2$  & \multirow{2}{*}{SDE} & \multirow{2}{*}{$O(d^{2}\times k^{2}\times n)$ } & \multirow{2}{*}{\texttt{ksimplex}} \tabularnewline
 &  & (with EKF) &  &  & \tabularnewline
\cline{1-6}  
  & \multirow{2}{*}{pMCMC} & \multirow{2}{*}{Particle Filter$^1$} & \multirow{2}{*}{Markov models} & \multirow{2}{*}{$O(1000\times k\times n)$ } & \multirow{2}{*}{\texttt{pmcmc}} \tabularnewline
Explore &  &  &  &  & \tabularnewline
\cline{2-6} 
$p(\theta|y)$ & \multirow{2}{*}{EK-MCMC} & SDE integration$^2$   & \multirow{2}{*}{SDE}  & \multirow{2}{*}{$O(k^{2}\times n)$ }& \multirow{2}{*}{\texttt{kmcmc}}\tabularnewline
 &  & (with EFK) &  &  & \tabularnewline
\hline 
\end{tabular}
\par\end{centering}
\end{adjustwidth}

\caption{\label{tab:algs}Algorithms implemented in PLoM, with corresponding characteristics.\protect \\
d: dimension of $\theta$, k: dimension of the system, n: number of observations.\protect \\
$1$: exact estimation of the likelihood, with Monte Carlo error\protect \\
$2$: deterministic approximation of the likelihood}
\end{table}

\vspace*{\fill}

 \pagebreak{}

\section{A simple illustration case: efficiently initialising a particle MCMC}

\begin{figure}[!h]
\begin{center}
\includegraphics[width=10cm]{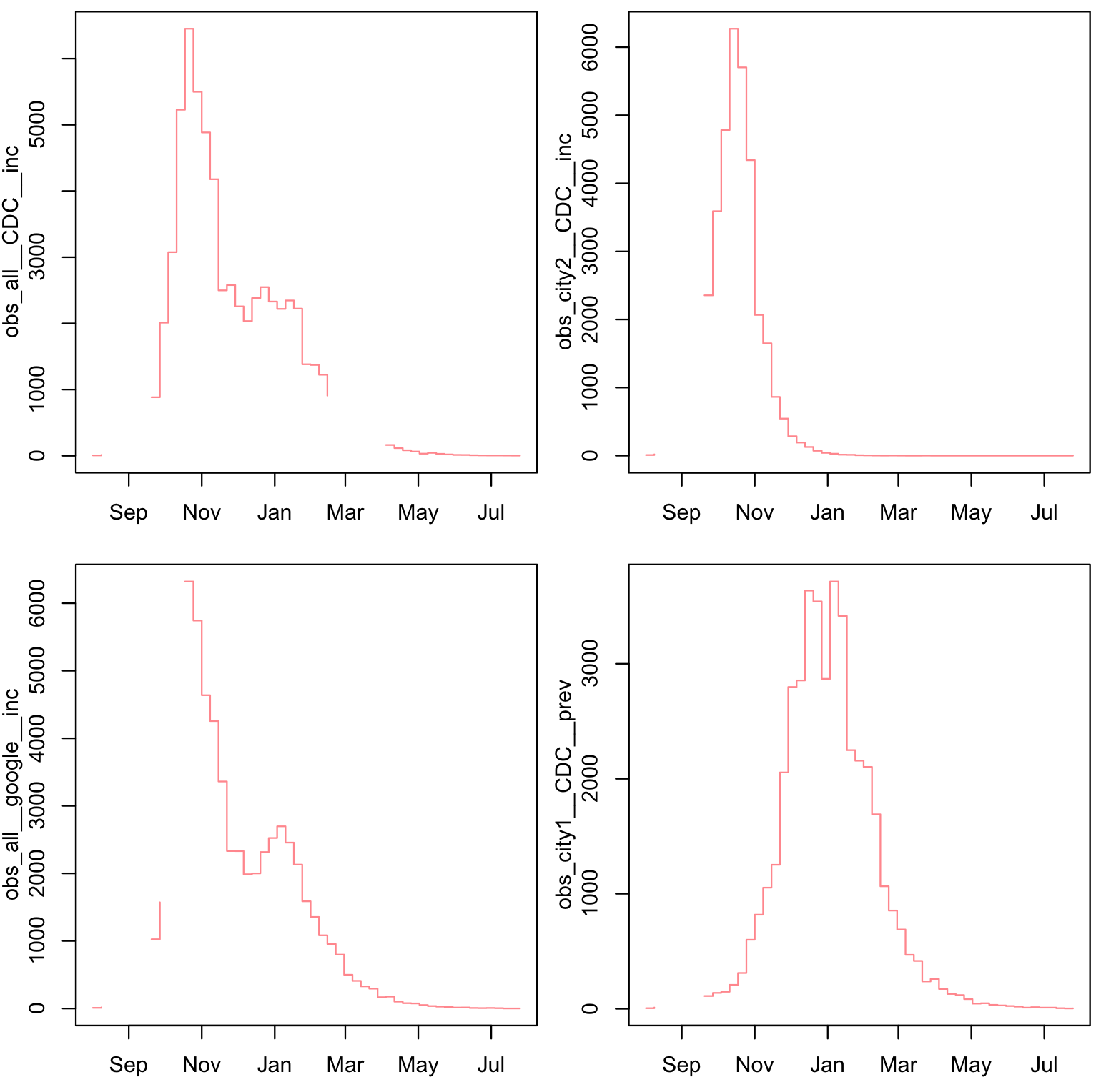}
\caption{\label{fig:Data}Simulated epidemiological data, reproducing the complexity of real datasets: both city-specific and global observations, different data streams, combination of prevalence and incidence observations, and missing data.}
\end{center}
\end{figure}

In this section, we will illustrate how the  \texttt{lhs}, \texttt{ksimplex} and \texttt{kmcmc} tools implemented in PLoM can be used to efficiently initialise a particle MCMC. We will work on a simple SI model, that can be found in the tutorials of the PLoM library. We incorporate environmental stochasticity by having the efficient reproduction number $R_0$ follow a diffusion, to capture its changes in time. We have generated data for two cities, in which the contact patterns (hence  $R_0$) may be different. The problem we will deal with is the estimation of the initial reproduction numbers in city 1 and 2, respectively $R_0^1(t_0)$  and $R_0^2(t_0)$, and the length $v$ of the infectious period, that can be considered to be the same in both cities. Observations are shown in Figure  \ref{fig:Data}, they consist of grouped incidence over cities 1 and 2, monitored by different sources (CDC and Google FluTrend for example), incidence in city 2, and prevalence in city 1. The process, contect and link files for this specific problem can be found in the examples library as \texttt{drift}, so that you can reproduce and play with the experiments presented here.

One way to obtain information regarding $R_0^1(t_0)$ , $R_0^2(t_0)$ and $v$ is to explore the density $p(\theta|y)$, with $\theta$ being $\{R_0^1(t_0),R_0^2(t_0),v\}$. As this cannot be done directly, it is common practice to use a Monte Carlo Markov Chain. Concretely, It is a process that travels around the $\theta$ space ($\mathbb{R}^3$ here). Even more concretely, suppose that at iteration $i$ the chain is in $\theta_i$: a new value $\theta^*$ is sampled from a multivariate gaussian density centered in $\theta_i$, which covariance will be noted $\Sigma_i$. If $p(\theta^*|y)p(\theta^*)$, that can be estimated in different ways as explained in Section 1,  is higher or not too much lower than $p(\theta_i|y)p(\theta_i)$ (there is a formula behind this), $\theta^*$ will be accepted and become $\theta_{i+1}$. If  $\theta^*$ is not accepted, the chain does not move and $\theta_{i+1}$ is equal to $\theta_{i}$. The theory says that when this chain converges, the sequence of samples it has generated can be treated as samples from $p(\theta|y)$ (see \cite{Andrieu2003} for more details). However, it is important to note that if a chain that has converged is run for $1000$ iterations, for example, the output is not equivalent to $1000$ independent and identically distributed samples from the target distribution. Again, there is a formula that gives you what is called the "Effective Sample Size" (ESS), the equivalent number of really independent samples you can consider to have obtained from your MCMC run. One last bit of theory: it has been shown that aiming for an acceptance rate around 23\% is good practice, in order to maximise the ESS \citep{Roberts1997}.

Getting back to practical considerations, what is needed to start the chain is an initial position $\theta_0$, and a covariance matrix $\Sigma_0$. In PLoM, if the pMCMC is launched from scratch, it will take the \texttt{guess} values as initial conditions, and the square of the \texttt{sd\_transf}'s (for standard deviation in the transformed space) that have been entered manually as diagonal terms for $\Sigma_0$. There are two things to note at this point: first of all this makes $\Sigma_0$ a diagonal matrix, which may be suboptimal if some components of $\theta$ are correlated (we will illustrate this later). Then, it is hard to guess what are good values for the \texttt{sd\_transf}'s as it would be optimal to set them close from the covariance of the posterior density, which is exactly what we don't know and are trying to figure out. We will now give an illustration of how things can go wrong or at least be inefficient if a pMCMC is poorly initialised.

\subsection{Examples of poor initialisations of the pMCMC}

\subsubsection{Starting from an arbitrary $\theta_0$}

\begin{figure}[!h]
\begin{center}
\includegraphics[width=13cm,trim=4 300 20 160, clip=true]{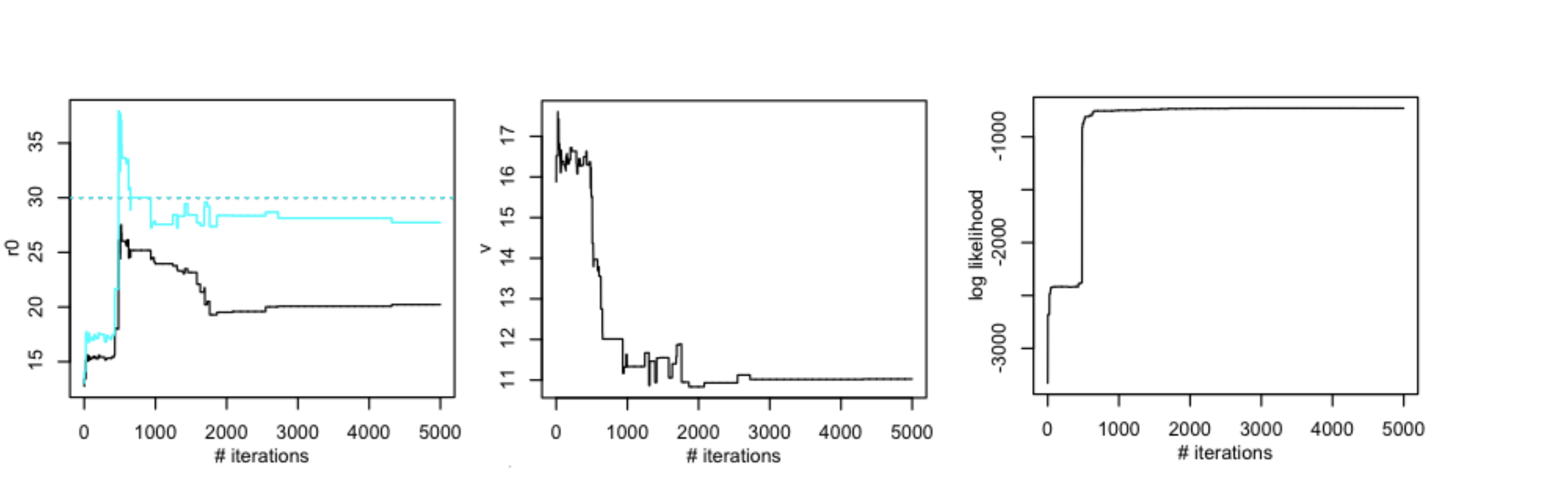}
\caption{\label{fig:WrongInit}Traceplots and corresponding outputs for a wrongly initialized pMCMC: more than 2000 iterations are lost to find the mode, albeit the model is fairly simple}
\end{center}
\end{figure}

\begin{figure}[!h]
\begin{center}
\includegraphics[width=13cm,trim=4 300 20 160, clip=true]{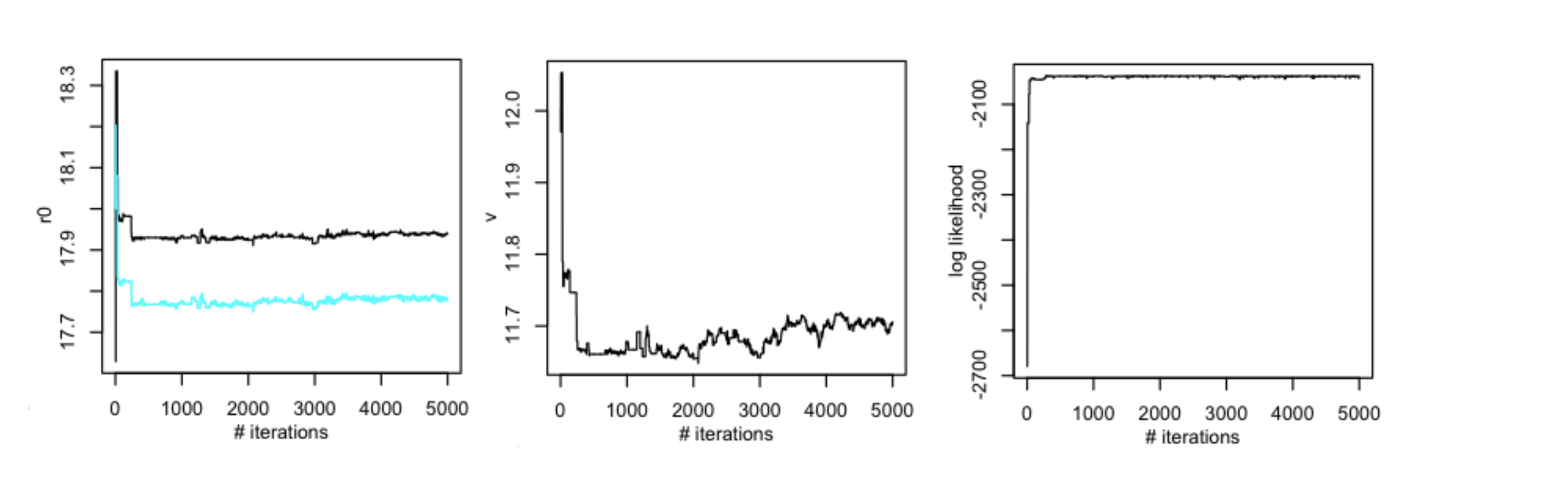}
\caption{\label{fig:WrongInit2}Traceplots and corresponding outputs for a wrongly initialized pMCMC: the chain stabilizes in a local mode where likelihood is below -2000 (the global mode is around -700). Once again, such problems can occur although the model is fairly simple.}
\end{center}
\end{figure}

To begin with, we will consider that the covariance matrix $\Sigma_0$ has been optimally chosen (case $d$ in Figure \ref{fig:Sigmas}), and that initial conditions for $R_0^1(t_0)$ , $R_0^2(t_0)$ and $v$ have arbitrarily been  set to 13, 13 and 16 respectively. The traceplot of one  corresponding output is shown in Figure  \ref{fig:WrongInit}: it takes more than 2000 iterations to find the mode, albeit the model used in this example is fairly simple. Another possible scenario is that the chain gets stuck in a local maximum, as shown in Figure \ref{fig:WrongInit2}. It does not mean that the pMCMC does not work, it is a simple consequence of the fact that the pMCMC is not an optimisation algorithm: it should be launched close from the mode (or at least from the highest modes, in a multimodal density).  We will show in the following subsection how such difficulties can be avoided.

\subsubsection{Using an arbitrary covariance matrix $\Sigma_0$}
We consider here the classic, non-adaptive version of the pMCMC algorithm: the same sampling covariance $\Sigma_0$ is used all along the chain. If  \texttt{sd\_transf}'s are set, for example, to 0.02 for $R_0^1(t_0)$ , $R_0^2(t_0)$ and $v$, and used to define $\Sigma_0$ as shown in case $a$ of Figure \ref{fig:Sigmas}, the algorithm performs very poorly. More than a million pMCMC iterations would be necessary to obtain the equivalent of 100 effectively independent samples. This can be translated into a measure of relative efficiency by comparing asymptotically the ESS of an MCMC with such $\Sigma_0$ to the ESS obtained with an optimal covariance $\Sigma_{opt}$. An estimate of the optimal covariance of reference is shown in Figure  \ref{fig:Sigmas}, case $d$; it has been empirically obtained from efficient and long runs of pMCMC on this dataset. Under this criteria, the arbitrarily chosen $\Sigma_0$ leads to a pMCMC that is 10 000 times  less efficient than what could be optimally done!

\[
Relative\; Efficiency(\Sigma)\;=\frac{ESS(\Sigma)}{ESS(\Sigma_{opt.})}
\]

\subsection{Improving the efficiency of the pMCMC by relying on the \texttt{lhs}, \texttt{ksimplex} and \texttt{kmcmc} algorithms}

\subsubsection{Using \texttt{lhs} and \texttt{ksimplex}  to determine $\theta_0$}

\begin{figure}[!h]
\begin{center}
\includegraphics[width=13cm,trim=4 250 20 240, clip=true]{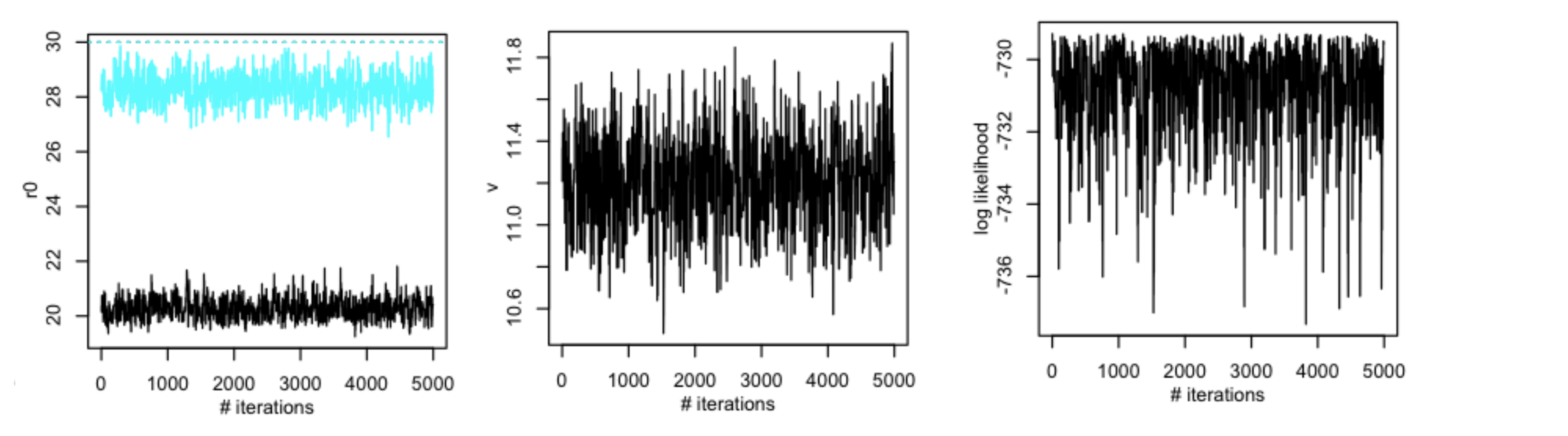}
\caption{\label{fig:GoodInit}Traceplots and corresponding outputs for a pMCMC with a $\theta_0$ adequately tuned with the lhs and ksimplex algorithm: no transient phase is observed, there are no  pMCMC iterations used for tuning}
\end{center}
\end{figure}

As indicated in Table \ref{tab:algs}, there are three options in PLoM to find an optimum: the \texttt{mif} algorithm, the \texttt{simplex} algorithm, and the \texttt{ksimplex} algorithm. All three could be used, but  \texttt{ksimplex} may be the wisest choice. The  \texttt{mif} algorithm does not use any approximation, but as it is based on a particle filter it may be computationally costly to run, while the goal here is only to find  a satisfactory approximation of the mode. The  \texttt{simplex} algorithm is the fastest of all three, as it only implies an ODE integration. However, it skips the stochastic components of the models, which can play an important role in models as the one we are looking at. Here, $R_0$ follows a diffusion to model its variations in time: considering a model without this diffusion makes it collapse to a model with constant $R_0$, which may lead to deceiving conclusions regarding the shape of the underlying likelihood function. Furthermore, $\theta$ can sometimes contain parameters that are only involved in the stochastic components of the model, as the volatilities of diffusing parameters, or the amplitude of environmental stochasticity. Running a deterministic simplex algorithm in such cases will only bring incomplete information. The ksimplex algorithm is both significantly faster than mif, and accounts for the stochastic components of the model. It can then be used to efficiently and reliably initialise a pMCMC algorithm. To avoid the risk of being stuck in a local maximum, the simplex algorithm should be associated with the lhs functionality of PLoM (for Latin Hyper-Square). It will randomly generate positions in the $\theta$-space, where an EKF will be run. The best position can be used as an initial condition for the optimisation algorithm simplex. The simplex generally converges quickly, and a pMCMC run from this position leads to the outputs shown in Figure  \ref{fig:GoodInit}, that can be favourably contrasted with Figures \ref{fig:WrongInit} and \ref{fig:WrongInit2}.

\subsubsection{Initialising the Adaptive pMCMC with an  EK-MCMC}

\begin{figure}[!h]
\begin{center}
\includegraphics[width=13cm,trim=4 220 20 110, clip=true]{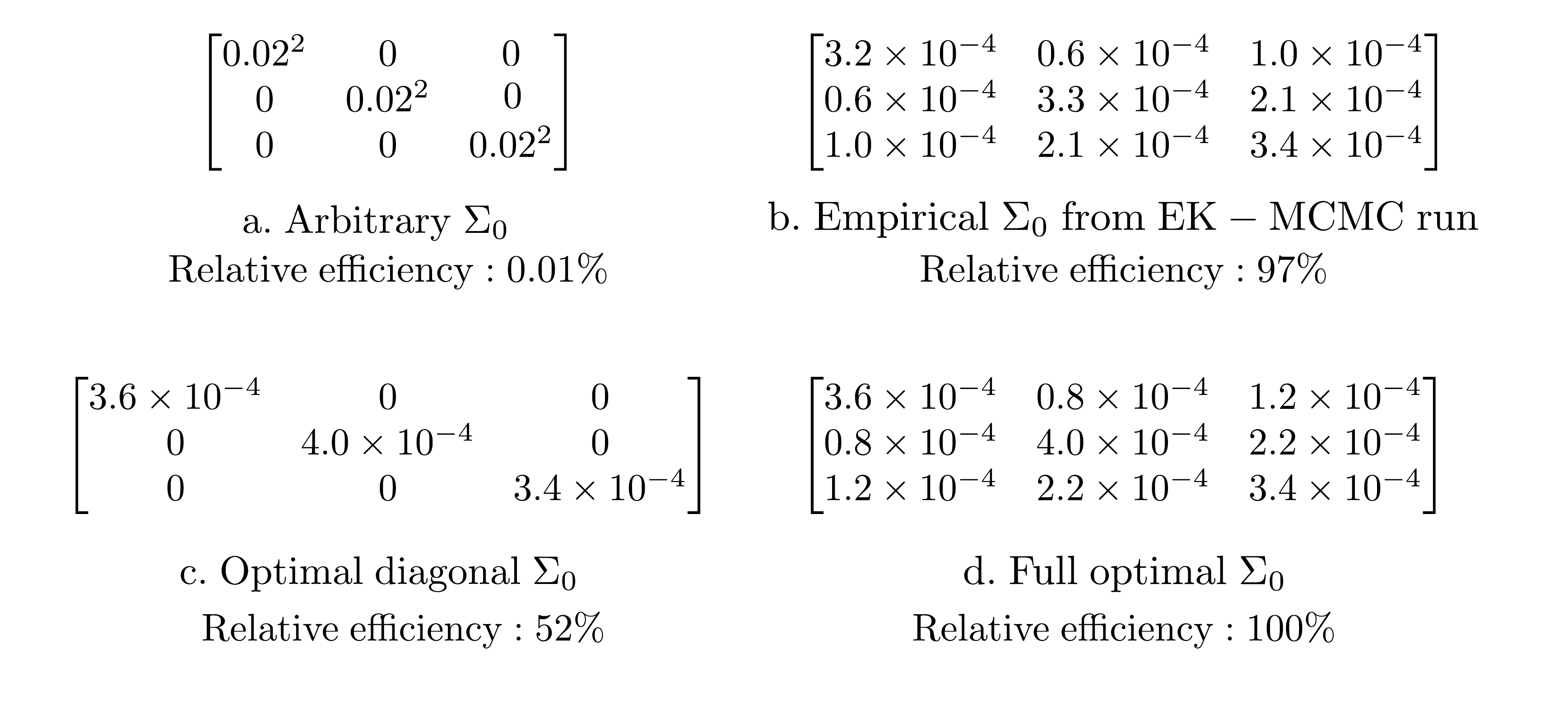}
\caption{\label{fig:Sigmas}Four experiments where an non-adaptive pMCMC has been run with different covariance matrices $\Sigma_0$. With each experiment, the sampling efficiency relative to the optimal one achieved in case $d$ is given.}
\end{center}
\end{figure}

The poor results of case $a$ in Figure  \ref{fig:Sigmas} show how inefficient can be a pMCMC algorithm based on arbitrarily chosen \texttt{sd\_transf}'s. It is based on the non-adaptive form of the pMCMC, introduced in \cite{Andrieu2010}.  By simply incorporating the adaptive principles presented in \cite{Roberts2009}, the covariance matrix $\Sigma_0$ can progressively be replaced by more suitable matrices $(\Sigma_i)$ empirically computed from previous samples of the run. This adaptation can significantly improve the pMCMC efficiency. However, the "learning" phase can be long, specially when $\Sigma_0$ is as inefficient as in case $a$ of Figure  \ref{fig:Sigmas}. We propose to tackle this task with a faster algorithm.

\begin{figure}[!h]
\begin{center}
\includegraphics[width=10cm]{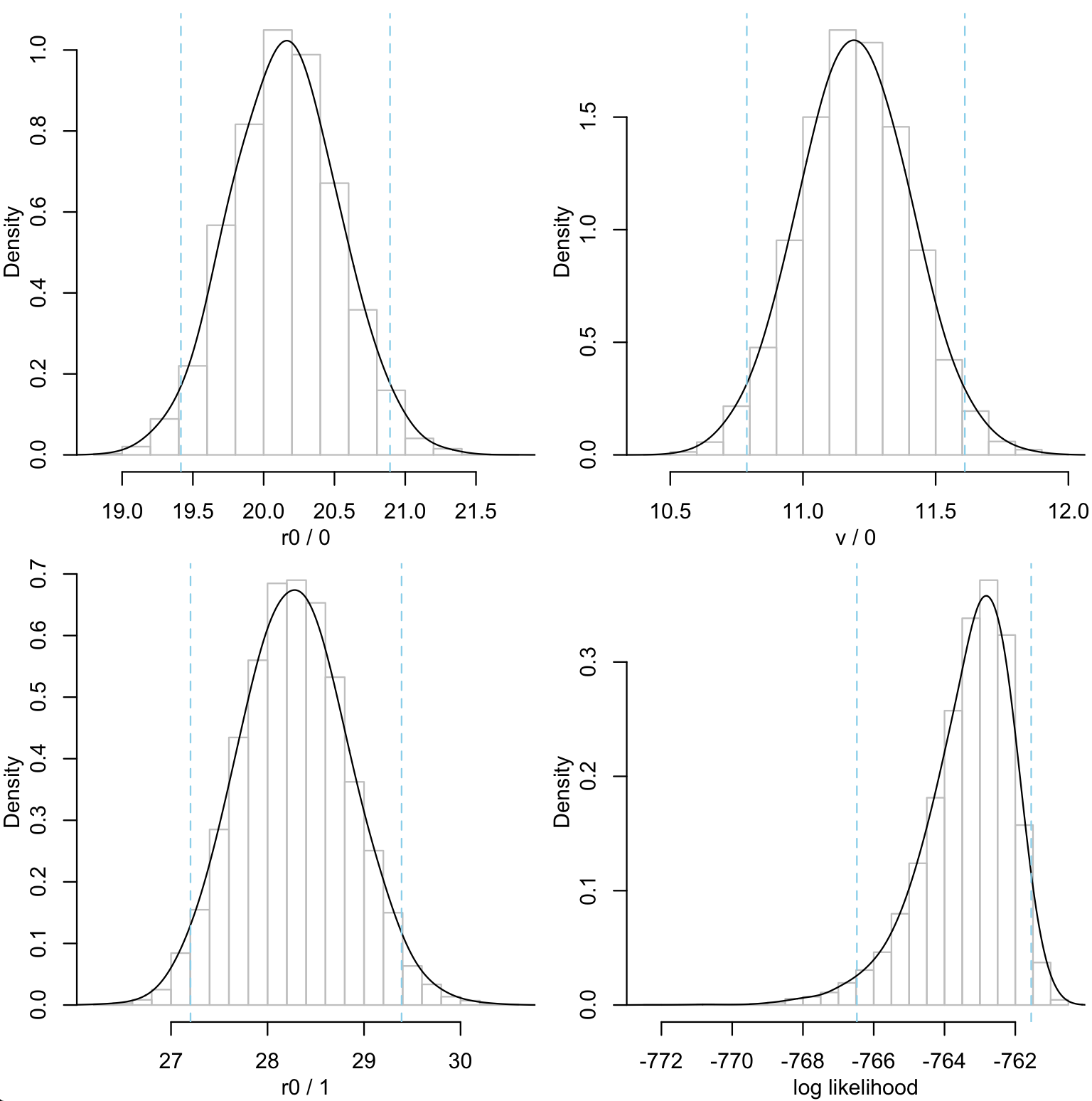}
\caption{\label{fig:Post}Posterior densities estimated from the outputs of an efficiently initialised pMCMC.}
\end{center}
\end{figure}

As shown in  Table \ref{tab:algs}, there is another algorithm implemented in PLoM to explore posterior densities: the EK-MCMC, that simply replaces the particle filter component of the pMCMC by an Extended Kalman Filter that quickly provides a good approximation of the likelihood. Case $b$ of Figure \ref{fig:Sigmas} indicates that by initialising $\Sigma_0$ with the empirical covariance matrix computed from the samples of an adaptive EK-MCMC (\texttt{kmcmc}) algorithm, the pMCMC will directly perform at 97\% of its nominal efficiency. Once again, there will be no need for a burn-in phase of the adaptive pMCMC if it has been adequately initialised using faster algorithms. The resulting posterior distributions shown in Figure  \ref{fig:Post} can then be obtained at a strongly reduced computational cost.

At last, case $c$ shows what would have been the efficiency of the pMCMC if by chance the \texttt{sd\_transf}'s had been set to their optimal values ($\Sigma_0$  still being a diagonal matrix). The absence of the extra-diagonal terms, in such a situation where parameters are correlated, significantly decreases the efficiency of the algorithm. By almost 50\% in this example.

 \pagebreak{}

\section{More details? Questions and answers}

\subsection{Similarly, can fast methods be used to reduce the computational cost of iterated filtering?}
By starting with a combination of the \texttt{lhs} and \texttt{ksimplex} methods, the iterated filtering algorithm should start close from the modes, making it converge much more quickly. 

\subsection{What about parallelisation?}
All particle-based methods (\texttt{smc},  \texttt{mif},  \texttt{pmcmc}) are automatically parallelised by distributing the particles over all available cores. This will work on your own machine as well as on clusters. In case you want to control the number of cores the method uses, you use can simply use the \texttt{-P} command. Parallelisation can substantially reduce the computing time it takes to run a particle filter!

Nevertheless, despite parallelisation the number of operations needed to run a particle filter stays the same: under this perspective the EKF still provides a substantial advantage in any case. Furthermore, there are cases where the covariance matrix of the EKF is sparse, which could make parallelisation profitable. This is the case, for example, if inference is done simultaneously on a variety of geographic patches in which epidemic dynamics are partly independent, as they only influence each others through migration. This is an open area for research and experimentation, that could lead to further parallelisation of the EKF and of the particle filters.

\subsection{What is the principle of the adaptive particle MCMC and its PLoM implementation?}
The adaptive pMCMC implemented in PLoM is inspired from the Adaptive Metrpolis algorithm introduced in \cite{Roberts2009}. It is based on the idea that as the chain progresses, the samples generated so far give a better and better idea of the structure of the posterior density, and should be used to tune the covariance matrix of the importance sampling distribution.

At the beginning of the chain, there are too few samples to reliably obtain a good estimate of the posterior density covariance. Thus, the particle MCMC (\texttt{pmcmc}) and EK-MCMC (\texttt{kmcmc}) algorithms implemented in PLoM work with two phases:\\

\begin{enumerate}
\item At iteration i, use the covariance $\Sigma_i = \epsilon_i^2\Sigma_0$, with $\epsilon_i$ defined in the following way:   
\begin{eqnarray*}
\epsilon_{0} & = & \frac{2.38^{2}}{dim(\theta)}\\
\epsilon_{i+1} & = & \epsilon_{i}\times\exp(a^{n}(AccRate-0.234))\}
\end{eqnarray*}
where $a$ is the cooling factor (can be controlled with the \texttt{-a} option of the \texttt{pmcmc} and \texttt{kmcmc} algorithms), and AccRate is the proportion of proposed samples $\theta^*$ that have been accepted up to this iteration. The  $\frac{2.38^{2}}{dim(\theta)}$  ratio has been shown to be the optimal scaling in a gaussian case \citep{Roberts1997}. 
\item After a certain number of samples have been accepted (controlled by the \texttt{-S} option),  $\Sigma_i$ is set to $\frac{2.38^{2}}{dim(\theta)}\times\Sigma_{Emp(1:i)}$, where $\Sigma_{Emp(1:i)}$ is the empirical covariance computed from the samples obtained up to iteration $i$.
\end{enumerate}
For more details, see  \cite{Roberts2009} and \cite{Dureau2012}.

\subsection{On practical examples, how much less computationally expensive is the EKF with regards to the particle filter?}
In \cite{Dureau2012}, two application examples of how the EK-MCMC can be used to initialise a pMCMC have been given. The first model is an SEIR model with a diffusing efficient contact rate $\beta_t$ and observed incidence, which leads to a state vector of five dimensions. Using a simple and direct implementation, the EKF is equivalent to propagating one vector of dimension $5(5+1)/2+5=20$ ($5(5+1)/2$ for half of the covariance matrix, that is symetric, and $5$ for the state vector). On the other hand, it has been determined that between 1000 and 2000 particles (vectors of dimension 5) are necessary to obtain satisfactory acceptance rates in the pMCMC. Hence, in this case between two and three orders of magnitude in terms of computational cost are gained when using the EKF.

In the second example, a slightly more complex SEIR model with two age classes is used, with both the children-to-children and adults-to-adults efficient contact rates diffusing, and both group-specific incidences being recorded. This leads to a state vector of dimension 10,  meaning EKF runs equivalent to propagating a vector of dimension 65, while 6000 where particles (of dimension 10)  considered necessary to run an acceptable particle filter. Once again, two to three orders of magnitude are gained when using the EKF.

\subsection{Are there alternatives to the particle MCMC?}
The $SMC^2$ algorithm (\cite{Chopin2011}) is an interesting alternative to the particle MCMC. The computational complexity appears to be equivalent, and the differences will lie in their implementation, how they can be parallelized on clusters, and how they can be combined with faster algorithms. Besides, an alternative to the adaptive pMCMC presented here would be a pMCMC where the MCMC component is no longer a random walk Metropolis algorithm, but rather an independent sampler efficiently defined from a preliminary EK-MCMC. This is just another example, there are globally many paths to explore!

\subsection{EKF can have numerical issues, how does PLoM deal with them?}
Practicionners know that nonlinear versions of the Kalman filters can have numerical issues that affect their stability. What has been done in PLoM is to fully exploit the adaptive integration tool of the GNU Scientific Library. It automatically adapts the integration time step to the magnitude of the variations induced by the dynamics of an ordinary differential equation, providing controlled numerical precision and higher stability. We take advantage of this approach by \emph{vectorizing} the deterministic equations followed by the covariance matrix in the EKF. By doing so, we turn the deterministic resolution of the SDE performed by the EKF into an ordinary differential equation, at least for the integration/prediction phases. This has improved the efficiency  and stability of the EKF. Of course, there may be more difficulties on the way, and other/better approaches
 to tackle the problem, we naturally welcome any comment or suggestion!

 \bibliography{Biblio}

\end{document}